\documentstyle[12pt,aaspp4]{article}

\begin{document}

\title{ASCA OBSERVATIONS OF THE IONIZED GAS IN THE SEYFERT GALAXY NGC~3783}

\author {I.M. George \altaffilmark{1, 2},
T.J. Turner \altaffilmark{1, 2}, 
R. Mushotzky \altaffilmark{1}, K. Nandra \altaffilmark{1, 3},
H. Netzer \altaffilmark{4}
}
\altaffiltext{1}{Laboratory for High Energy Astrophysics, Code 660,
	NASA/Goddard Space Flight Center,
  	Greenbelt, MD 20771}
\altaffiltext{2}{Universities Space Research Association}
\altaffiltext{3}{NAS/NRC Research Associate}
\altaffiltext{4}{School of Physics and Astronomy and the Wise Observatory,
        The Beverly and Ramond Sackler Faculty of Exact Sciences,
        Tel Aviv University, Tel Aviv 69978, Israel.}

\slugcomment{Resubmitted to {\em The Astrophysical Journal} (manuscript 37708)}

\begin{abstract}
We present the results from {\it ASCA} observations of NGC~3783
carried out during 1993 and 1996. 
Variability is observed both between the two epochs and within the individual 
observations, the latter due to a non-stationary process.

We find the spectra at both epochs to contain features due to absorption by 
ionized material, predominantly due to O{\sc vii} and O{\sc viii}.
We find the opacity for such material to decrease by a factor $\sim$2 in 
the $\sim 0.7$--1.0~keV band between the epochs while the photoionizing 
flux increases by $\sim$25\%.  
By comparison with detailed photoionization models we show this behaviour is 
inconsistent with that expected from a single-zone of photoionized gas. 
A deficit in the data compared to such models, consistent with excess 
absorption by O{\sc viii}, supports the suggestion that the material 
consists of two or more zones

Significant Fe $K$-shell emission is also observed during both epochs.
We show this emission is dominated by the asymmetric line profile expected 
from the innermost regions of the accretion flow.
We find no evidence that the Fe emission varied in either shape or equivalent 
width between the two epochs.
\end{abstract}



\clearpage
\section{INTRODUCTION}
\label{Sec:intro}

The almost face-on spiral (SBa) galaxy NGC~3783 ($z = 0.0097$) is one of the 
brighter Seyfert galaxies in the sky in all wavebands 
(e.g. Alloin et al. 1995 and references therein). 
At X-ray wavelengths, NGC 3783 was first detected in the {\it Ariel-V}
sky-survey and has subsequently been observed by all major X-ray instruments 
(e.g. Malizia et al 1997 and references therein).
Evidence for deep absorption features in the 0.5--1.5~keV band was first 
obtained in a {\it ROSAT}\ observation (Turner et al. 1993) and 
confirmed in a subsequent {\it ASCA}\ observation 
(George, Turner \& Netzer 1995, hereafter G95).
The latter workers demonstrated the opacity was dominated by $K$-shell 
absorption edges due to O{\sc vii} and O{\sc viii} within a screen of 
material (with an effective hydrogen column density 
$\sim 10^{22}\ {\rm cm^{-2}}$) within the column--of--sight to the central 
source of X-ray emission.
Subsequent {\it ASCA}\ observations have revealed such screens of photoionized 
material (so--called  "warm absorbers'') to be a common features in Seyfert 
galaxies (Reynolds 1997; George et al 1998a, hereafter G98).
G95 also suggested the tentative detection of the emission predicted to arise 
from within such a screen of ionized gas if it subtends a significant solid 
angle at the central source.

Here we present the results from the analysis of four new {\it ASCA}\ 
observations of NGC~3783 performed in 1996 July, along with a 
re-analysis of the two previous observations carried out in 1993 December
(and as reported previously by G95, Reynolds 1997 and G98).
The observations performed in 1996 were each separated by $\sim$2 days, with 
the hope (based on the behaviour seen during 1993) of measuring the reaction 
of the ionized screen to changes in the illuminating continuum, and hence 
better constrain the location of and physical conditions within the screen.
Unfortunately, the photoionizing source did not co-operate, exhibiting
only relatively low-amplitude variations during the 1996 observations.
However, we as we shall show, useful insights regarding the nature of 
the ionized, circumnuclear material can still be obtained by comparision 
of the 1996 and 1993 observations.

The observations and data-screening employed are described in 
\S\ref{Sec:data_anal}. 
In \S\ref{Sec:spatial} we describe the results of our spatial and temporal 
analysis.
The results of our spectral analysis of the absorption due to the ionized gas 
are presented in \S\ref{Sec:spectral-cont}, confirming the presence of the 
ionized material during both 1993 and 1996. However we find the opacity 
of the ionized material to decrease by a factor $\sim$2 between these
epochs, and discuss physically realistic explanations.
In \S\ref{Sec:nldl} we discuss the constraints which can be placed on 
the emission due to Fe $K$-shell processes in the 5--7~keV band.
Finally, in \S\ref{Sec:Discuss} we review our findings in the context of the 
results from similar objects and future observations.

\section{OBSERVATIONS AND DATA REDUCTION}
\label{Sec:data_anal}

{\it ASCA} consists of four identical, co-aligned X-ray telescopes 
(XRTs; Serlemitsos et al. 1995). Two solid-state imaging spectrometers 
(known as SIS0 and SIS1), each consisting of four CCD chips, sit at the 
focus of two of the XRTs, and provide coverage over the
$\sim$0.4--10~keV band (Burke et al. 1994). Two gas imaging spectrometers
(known as GIS2 and GIS3) sit at the focus at the focus of the other two
XRTs, and provide coverage over the $\sim$0.8--10~keV band (Ohashi et al.
1996).
Further details on the satellite, its instrumentation and performance can be 
found in Makishima et al. (1996) and references therein.

The four new observations of NGC~3783 reported here were carried out over the 
period 1996 July 09--16. 
We have also performed a re-analysis of the 2 observations of NGC~3783 
carried out in 1993 December 19--23, which have been previously described in
G95, Reynolds (1997) and G98. 
However, here we utilize raw data from the {\tt Rev2} 
processing\footnote{hence raw data from the same processing configuration 
(6.4.2) is used for all the datasets}, 
along with new screening criteria and the latest calibration files.
These changes result in slightly different values for some parameters
compared to previously published results.
The full observing log is given in Table~1.

The data-selection and screening criteria used here are similar to those 
described in Nandra et al (1997a).
We started from the 'unscreened' event files.
Data collected in {\tt FAINT} and {\tt BRIGHT} data modes were combined, as 
were data obtained during all three telemetery modes.
These data were then screened\footnote{the '{\tt mkf}-file', describing the 
orbital parameters of the satellite, was corrected for 
the errors in the version initially used during the pre-processing of the 
1996 datasets}
using the {\tt ASCASCREEN}/{\tt XSELECT} script (v0.43) within the 
{\tt FTOOLS} package (v3.6) such as only to include data collected 
when:
the spacecraft was outside of the South Atlantic Anomaly; 
the radiation belt monitor rate was less than $500\ {\rm ct\ s^{-1}}$; 
magnetic cut-off rigidity was $>$6 GeV/c; 
angular offset from the the nominal pointing position was $<$36 arcsec;
elevation angle above the Earth's limb was $>5^{\circ}$ for GIS data, 
	and $>10^{\circ}$ for SIS data.
In the case of SIS data (only), the following additional 
criteria were applied:
that the Bright Earth angle (elevation angle above the sun-illuminated Earth's 
	limb) was $>20^{\circ}$; 
that the spacecraft was at least 100 s beyond the day/night terminator;
that the CCD pixel threshold was $<$100 ($<$75 for the 1993 observations).
'Hot' and 'flickering' pixels were removed from the SIS using the standard
algorithm. Only SIS 'grades' 0, 2, 3 and 4 were included in the  analysis.
The original pulse-height assignment for each event was converted to 
a pulse-invariant (PI) scale using {\tt SISPI} (v1.1).
In the case of the GIS data (only), the recently-discovered method of 
rejecting 'hard particle flares' using the so-called {\tt HO2} count rate
was employed, as was the standard 'rise-time' criteria.
It should be noted that these screening criteria are more conservative than
those used in G95 and resulted in lower effective exposure times of
$\sim$17~ks and $\sim$15~ks in each instrument for the Dec 19 and Dec 23
observations (respectively). 

\section{SPATIAL \& TEMPORAL ANALYSIS}
\label{Sec:spatial}

Images were extracted for each instrument during each observation.
In all cases, a bright source was detected with an 
X-ray centroid consistent
with the optical position of NGC~3783 to within the uncertainty in the
positional accuracy of the {\it ASCA} attitude reconstruction 
(90\% confidence radius of 40 arcsec, Gotthelf \& Ishibashi 1997).

No serendipitous sources were detected which are likely to contaminate 
significantly the analysis of the target. 
Analysis of the archival {\it ROSAT} observation (that made in 1993 June)
using the Position Sensitive Proportional Counter (PSPC) in the focal 
plane reveals no evidence for extended emission.

\subsection{Extraction Cells}
\label{Sec:spatial-xtractcells}

Extraction cells were defined for the subsequent temporal and spectral 
analysis. 
In the case of the SIS data, the source region employed was circular of radius 
$\sim 3.2$~arcmin centered on NGC~3783. 
From the nominal point-spread function (psf) of the XRT/SIS instrument,
such a region contains $\sim84$\% of the total source counts.
For the count rate and spectrum of NGC~3783, we find the background 
(instrumental and diffuse X-ray background) within the source region 
to be $\gtrsim 10$\% of the observed source flux $\gtrsim 5$~keV.
Thus an extraction cell was defined to provide an estimate of the background 
for each SIS detector which consisted of the whole CCD chip excluding a 
circular region of $\sim 4.3$~arcmin centered on the source.
In the case of the GIS data, the source region was circular of radius
$\sim 5.2$~arcmin centered on NGC~3783.
From the psf of the XRT/GIS instrument such a region contains $\sim89$\% of 
the total source counts.
The GIS has a higher background than the SIS, contributing $\gtrsim 10$\% of 
the observed source flux $\lesssim 1$~keV and $\gtrsim 4$~keV.
Thus an annulus, centered on the source and covering $\sim 5.2$--9.8~arcmin
was used to provide an estimate of the
background for each GIS detector.
All fluxes and luminosities quoted below ({\it not} count rates) have been 
corrected for the fraction of the source photons falling outside the source 
extraction cells and for the contamination of source counts in the background 
extraction cells ($\sim 6$\% and $\sim 11$\% of the source counts for 
the XRT/SIS and XRT/GIS instruments respectively).

\subsection{Variability Studies}
\label{Sec:temporal}

Light curves were constructed for the source and background regions for 
several different energy ranges for each detector during each observation.
To increase the signal-to-noise ratio, the light curves from SIS0 and SIS1 
and from GIS2 and GIS3 were combined in each case.
The light curves were then rebinned on a variety of timescales.

\begin{figure}
\plotfiddle{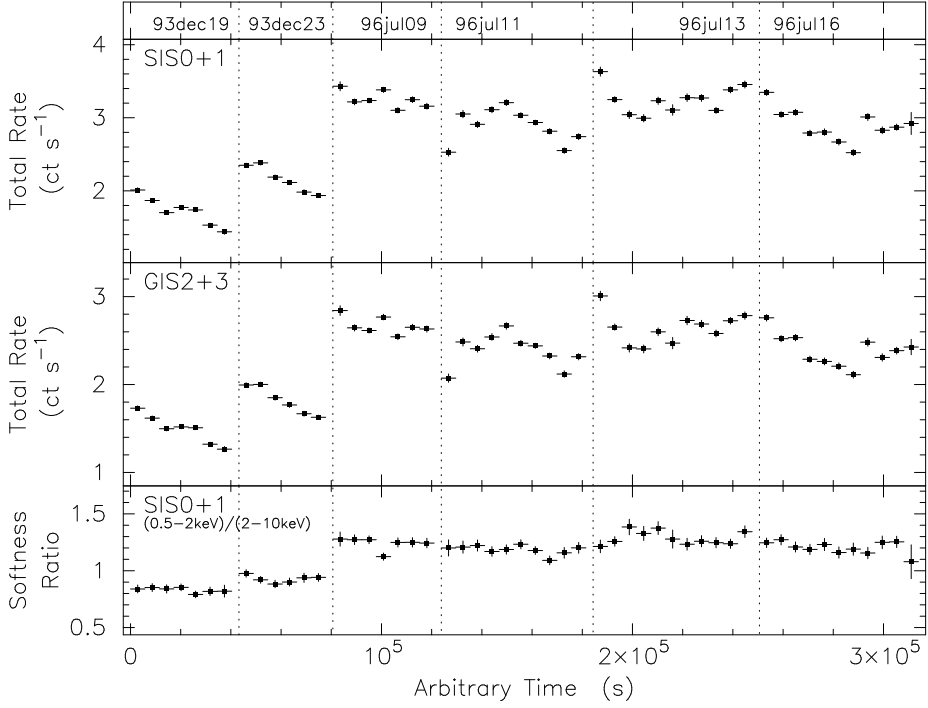}{10cm}{270}{50}{50}{-225}{300}
\caption{Light curves for the six {\it ASCA}\ observations 
of NGC~3783 reported here, employing a bin size of 
5760~s.
The upper two panels show the summed light curves obtained
for SIS and GIS detectors, whilst the lower 
panel shows the softness ratio (the ratio of the summed count rates observed 
in the 0.5--2~keV and 2--10~keV bands) for the SIS datectors.
In all cases, the y-axis is scaled to cover a factor of 0.4 to 1.6 
of the mean.
Significant variability is clearly apparent in the observed count rates 
both within the individual observations, and between the 1993 and 1996 
epochs.
However, whilst there is a clear change in the softness ratio between 
the 1993 and 1996 observations, there is no evidence for significant 
variability in this quantity within individual observations.
\label{fig:lc5760}}
\end{figure}

In Fig.~\ref{fig:lc5760} we show the light curves for the 0.5--10~keV SIS and 
2--10~keV GIS with a bin size of 5760~s (approximated the obital period of 
the {\it ASCA}).
Variability is apparent both between and within individual observations.
From Fig.~\ref{fig:lc5760} and Table~1 it can be seen that 
the mean count rate during the 1996 observations is a factor $\sim 1.5$ 
higher than during the 1993 observations.
The softness ratio, shown in the lower panel of Fig.~\ref{fig:lc5760},
exhibited a clear increase between the 1993 and 1996 observations.
As we shall show below, this primarily due to a decrease in the opacity of 
ionized material between the two epochs.

Interestingly NGC~3783 is seen to exhibit significant variability during in 
all but one (Obs. c) of the observations.
In Table~2 we list the normalised 'excess variance', 
$\sigma^2_{rms}$, of each light curve using the 
prescription\footnote{We note there is a typographical error in the 
expression for the error on $\sigma^2_{rms}$ given in Nandra et al (1997a), 
whereby the expression within the summation should be squared.}
given in Nandra et al (1997a).
We find no significant differences between $\sigma^2_{rms}$ from the light 
curves constructed in the hard (2--10~keV) and soft (0.5--2~keV)
bands.
This is confirmed by the lack of significant intra-observation variability 
seen in the softness ratio.
However, there are clear variations in $\sigma^2_{rms}$ from observation to 
observation. For all four bands we find $\sigma^2_{rms}$ inconsistent with a 
constant at $>$99\% confidence, indicating the process responsible for the 
variability is not stationary.

\section{SPECTRAL ANALYSIS EXCLUDING THE IRON $K$-SHELL EMISSION}
\label{Sec:spectral-cont}

Given the lack of large-amplitude variability within each observation
(Fig.~\ref{fig:lc5760}), here we consider the mean spectra for 
each observation. 
Source and background spectra were extracted from the cleaned event 
list of each detector using the extraction cells described in
\S\ref{Sec:spatial-xtractcells}. 
For the SIS datasets, appropriate detector redistibution matrices were 
generated using {\tt sisrmg} (v1.1). For the GIS datasets the 
redistibution matrices released on 1995 Mar 06 (generated by
{\tt gisres} v4.0) were used. 
Ancillary response files were generated for all detectors using 
{\tt ascaarf} (v2.72).

In all cases described below, the spectral analysis is performed on the
data from all four instruments simultaneously, with different relative
normalizations to account for (small) uncertainties in the determination of
their effective areas. 
(The correction for the fraction of the source counts falling outside the 
source region used for each detector is corrected for within {\tt ascaarf}.)
Data below 0.6 keV were excluded from the spectral analysis as it is commonly 
accepted that there are uncertainties associated with the calibration of the 
XRT/SIS system below this energy. 
However, whilst the calibration is suspect at these energies, it is considered 
unlikely to be in error by $\gtrsim20$\%. 

The individual spectra were grouped such as to contain a minimum of 20
counts per new bin, and hence allowing $\chi^2$ minimization techniques to
be employed within the {\tt XSPEC} (v10.00) spectral analysis package. 
For all the analysis presented in this section, we have assumed spectral
models consisting of an underlying power-law continuum (of photon index
$\Gamma$) absorbed by a neutral column
($N_{H,0}$) at zero redshift. We assume a Galactic column density  of
$N_{H,0}^{gal} = 8.7\times10^{20}\ {\rm cm^{-2}}$ ($E(B-V)=0.12$ for a
Galactic gas--to--dust ratio) along the column--of-sight to NGC~3783, as
derived from the fits of the {\it HST} FOS spectrum (Allion et al.\ 1995),
and consistent with the 21~cm measurements in this direction (Stark et al.
1992). Unless stated otherwise, we fix $N_{H,0}= N_{H,0}^{gal}$ in the 
spectral analysis presented below.

As will be discussed in \S\ref{Sec:nldl}, a strong emission line is present 
in the spectrum of NGC~3783 during all observations as a result of 
Fe $K$-shell fluorescence.
The best-fitting parameters of such a line are dependent upon the form of the 
underlying continuum, which itself is highly correlated with the absorption 
present in the soft X-ray band. 
In this section, we exclude the 5--7~keV band (source frame) from our 
analysis  and concentrate on the form of the continuum and the nature of the 
absorber. 
The characteristics of the emission line and its (small) effect on the 
properties of the continuum and absorption are postponed to 
\S\ref{Sec:nldl}. 

\subsection{The ION models}
\label{Sec:ion}

Strong absorption features are clearly evident within the 0.6--3~keV band of 
all six datasets.
Models including photoionized material within the column--of--sight to 
NGC~3783 have been suggested previously to offer the most viable explanation 
for such features in this and other Seyfert 1 galaxies.
Thus we start by considering such a model, based on detailed photoionization 
calculations.

We employ the same photoionization calculations used by G98. Model 
spectra were generated using the photoionization code {\tt ION} 
(version {\tt ION96}) to calculate the physical state of a slab of gas when 
illuminated by an ionizing continuum (Netzer 1993, 1996).
The code includes all 
important excitation and ionization processes, full temperature and radiative 
transfer solutions, and emission, absorption and reflection by the gas are 
calculated self-consistently assuming thermal and ionization equilibrium. 
We assume an illuminating continuum typical of AGN of the luminosity of 
NGC~3783 (the 'weak IR' case of Netzer 1996, except for the spectral index in 
the X-ray band), solar abundances and a density of 
$n = 10^{11}\ {\rm cm^{-3}}$. Following Netzer (1996) and G98, 
the intensity of the ionizing continuum is parameterized by the 
'X-ray ionization parameter', $U_X$, defined as 
\begin{equation}
\label{eqn:U_X}
U_X = \int^{10\ {\rm keV}}_{0.1\ {\rm keV}}
\frac{Q(E)}{4 \pi r^2 n c} dE
\end{equation}
where $Q(E)$ is the rate of photon emission at energy $E$, $r$ the distance 
from the source to the illuminated gas. Further details on the assumptions 
made in the {\tt ION} models, the method by which they were included in the 
spectral analysis, and conversion factors between $U_X$ and 
ionization parameters defined over the entire photoionizing 
continuum ($>$13.6~eV) for various spectral forms can be found in G98.

\subsection{Results assuming a single-zone ionized-absorber}
\label{Sec:abs_only}

\begin{figure}
\plotfiddle{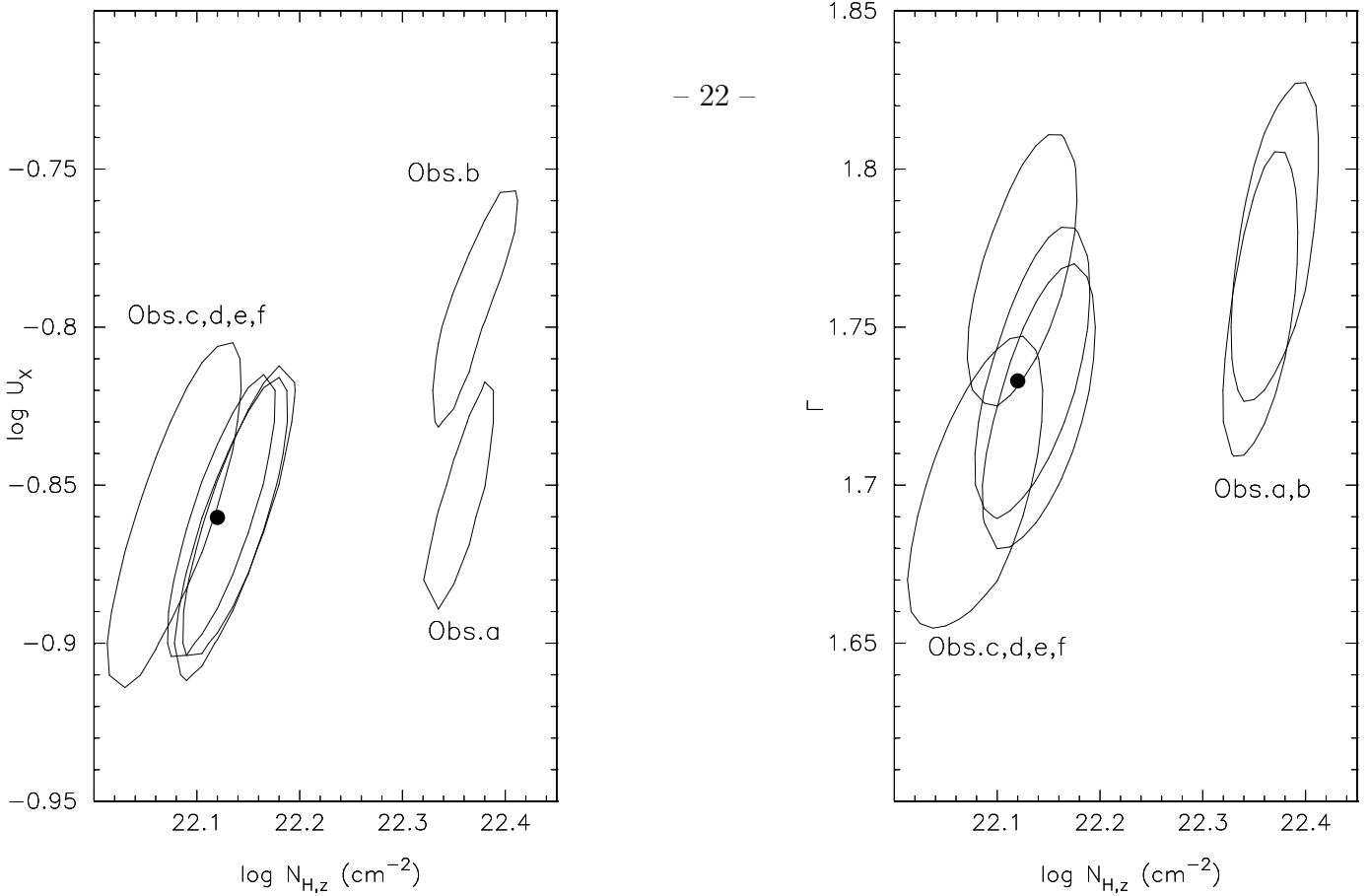}{10cm}{270}{50}{50}{-225}{-200}
\caption{Contours showing the 90\% confidence regions (for 3 interesting 
parameters) in the $N_{H,z}$--$U_X$ and $N_{H,z}$--$\Gamma$ planes 
for each of the observations of NGC~3783
assuming a model consisting of an underlying powerlaw continuum, 
a single-zone ionized absorber and with $N_{H,0}=N_{H,0}^{gal}$.
As also apparent from Table~3, 
all four datasets obtained in 1996 (Obs.c -- f) are consistent with 
mean values of
$< N_{H,z} >_{96} = 1.32\times10^{22}\ {\rm cm^{-2}}$,
$< U_X >_{96} = 0.138$,
and $< \Gamma >_{96} = 1.73$
(donoted by the filled circle).
A larger opacity is evident in the two datasets obtained 
in 1994 (Obs.a \& b), although it should be noted 
that the assumed model does not provide 
a statisfactory description in these cases
(see \S\ref{Sec:abs_only_93}).
\label{fig:ion_ngal_contours}}
\end{figure}

Assuming a single powerlaw continuum, absorbed by a screen of ionized 
material at the redshift of the source ($N_{H,z}$), and by a screen of 
neutral material at $z=0$ ($N_{H,0}$), in all cases we found $N_{H,0}$ 
consistent with $N_{H,0}^{gal}$.
Thus the analysis was repeated with $N_{H,0}$ fixed at this value.
The results are listed in Table~3,
and the contours of the 90\% confidence regions in the 
$N_{H,z}$--$U_X$ and $N_{H,z}$--$\Gamma$ planes are shown 
in Fig~\ref{fig:ion_ngal_contours}.

For reasons which will become apparent, we first consider the four datasets 
taken during the 1996 epoch, then compare these results to those from the 
datasets obtained during 1993.

\subsubsection{The 1996 observations}
\label{Sec:abs_only_96}

From Table~3, it can be seen that the single-zone 
photoionization model provide fits which are deemed acceptable 
for all 4 datasets obtained during the 1996 epoch
(with $P < 0.95$ and $0.8 \leq \overline{R_{0.6}} \leq 1.2$, see table notes).
Indeed, as is clearly demonstrated in
Fig~\ref{fig:ion_ngal_contours}, the best-fitting parameters are 
all consisent with an underlying continuum with $\Gamma \simeq 1.74$,
absorbed by a screen of ionized material with 
$N_{H,z} \simeq 1.3\times10^{22}\ {\rm cm^{-2}}$
and
$U_{X} \simeq 0.14$.
The intensity of the photoionizing source does exhibit statistically
significant variations between the four observations. However 
these are of relatively low amplitude
(Fig.~\ref{fig:lc5760}).
For example, the implied luminosity over the 0.5--2.0~keV band, $L_{0.5-2}$
has a mean value 
$<L_{0.5-2}>_{96} = 2.07\times10^{43}\ {\rm erg\ s^{-1}}$
during 1996, and exhibits variability $\lesssim 10$\%
between the observations
(Table~3).

\begin{figure}
\plotfiddle{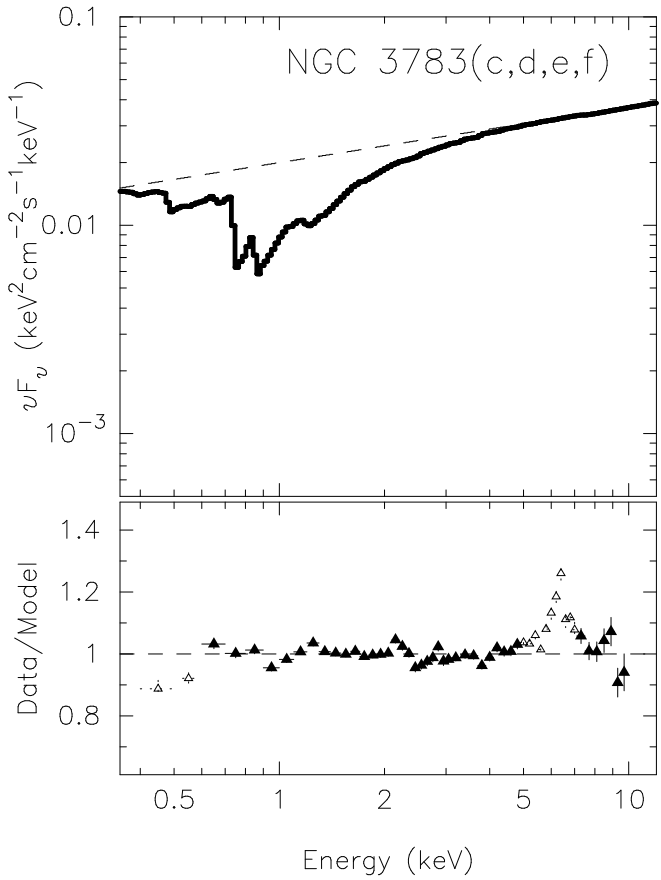}{10cm}{0}{50}{50}{10}{-200}
\caption{The upper panel shows (in bold)
the best-fitting, single-zone 
photoionization model obtained from simultaneous spectral 
analysis of the {\it ASCA} observations obtained during 1996 (see
\S\ref{Sec:abs_only_96}),
along with (dashed) the implied underlying continuum. 
Both curves have been corrected for Galactic absorption
($N_{HI}^{Gal} = 8.7\times10^{20}\ {\rm cm^{-2}}$).
The model represents a powerlaw continuum of photon index 
$< \Gamma >_{96} = 1.73$ absorbed by a screen of ionized material 
within the column--of--sight of column density 
$< N_{H,z} >_{96} = 1.32\times10^{22}\ {\rm cm^{-2}}$
and ionization parameter
$< U_X >_{96} = 0.138$.
The lower panel shows the mean data/model ratios.
The filled triangles show the (error-weighted) means of the
ratios from the individual detectors for the energy-bands used 
in the spectral analysis, rebinned in energy-space for clarity.
The open triangles show the corresponding rebinned, mean ratios when the
best-fitting model is extrapolated $<0.6$~keV and into the 5--7~keV band.
As noted in \S\ref{Sec:abs_only_96}, there are four major discrepancies 
between the data and model: 
the Fe $K$-shell emission (discussed in \S\ref{Sec:nldl}),
the instrument features $< 0.6$~keV and in the 2--3~keV band,
and the '1~keV deficit' (discussed in \S\ref{Sec:93vs96}).
\label{fig:all96_ion08_ngal_noline_ufmodelratio}}
\end{figure}

We have repeated the spectral analysis, but simultaneously fitting 
the data from all four detectors and all four observations obtained during 
1996, again allowing the relative normalizations to vary. 
We find best-fitting values of
$< N_{H,z} >_{96} = 1.32^{+0.07}_{-0.06}\times10^{22}\ {\rm cm^{-2}}$,
$< U_X >_{96} = 0.138^{+0.006}_{-0.006}$,
and $< \Gamma >_{96} = 1.73^{+0.02}_{-0.01}$, in
excellent agreement with the values anticipated from 
Table~3.
(For reference, for the assumed XUV spectrum, $< U_X >_{96}$ 
corresponds to an ionization parameter integrated over all 
energies $>$1~Ryd of $U = 19.5$.)
The spectrum implied by these observations is shown in 
the upper panel of Fig.~\ref{fig:all96_ion08_ngal_noline_ufmodelratio}.
From the mean data/model ratios 
shown in the lower panel of Fig.~\ref{fig:all96_ion08_ngal_noline_ufmodelratio},
noticable discrepancies between the data and model occur at 
four locations within the spectrum.
The most significant feature --- an excess of observed counts compared to
        the model in the 5--7~keV band --- is of course that expected 
	due to Fe $K$-shell emission, and is discussd in 
	\S\ref{Sec:nldl}.
Two additional features are also anticipated and due to 
	instrumental discrepancies. 
These are the deficit of observed counts $< 0.6$~keV (with 
	an amplitude data/model$\overline{R_{0.6}} \sim 0.9$) consistent with
	accepted inaccuracies in the calibration of the XRT/SIS instrument 
	at these energies, 
and the feature in the 2--3~keV band 
	due to slight errors in the modelling of the $M$-shell features 
	in the Au coating the XRTs.
The final discrepancy is  the 
	deficit of observed counts compared to the model in the 
	0.8--1~keV band and 
has an astrophysically-interesting interpretation, lying in a spectral band 
where there are no known problems with the calibrations of the instruments.
We postpone further discussion of this '1~keV deficit' to \S\ref{Sec:93vs96}.

To summarize, the basic conclusion is that over the week of the 1996 
observations, the underlying continuum in NGC~3783 was consistent with a
constant spectral index of $< \Gamma >_{96}$ and exhibited only low-amplitude 
variations in X-ray luminosity.
The bulk of the opacity evident during these observations is due to the 
column--of--sight to the central source being covered by a screen of ionized 
material of constant column density ($< N_{H,z} >_{96}$), density and radius 
(giving rise to $< U_X >_{96}$).
The lack of large-amplitude variations in the underlying continuum, combined 
with the achievement of statistically-acceptable fits using the {\tt ION} 
models, indicates the ionized material is consistent with being approximately 
in ionization equilibrium.

\subsubsection{The 1993 observations}
\label{Sec:abs_only_93}

\begin{figure}
\plotfiddle{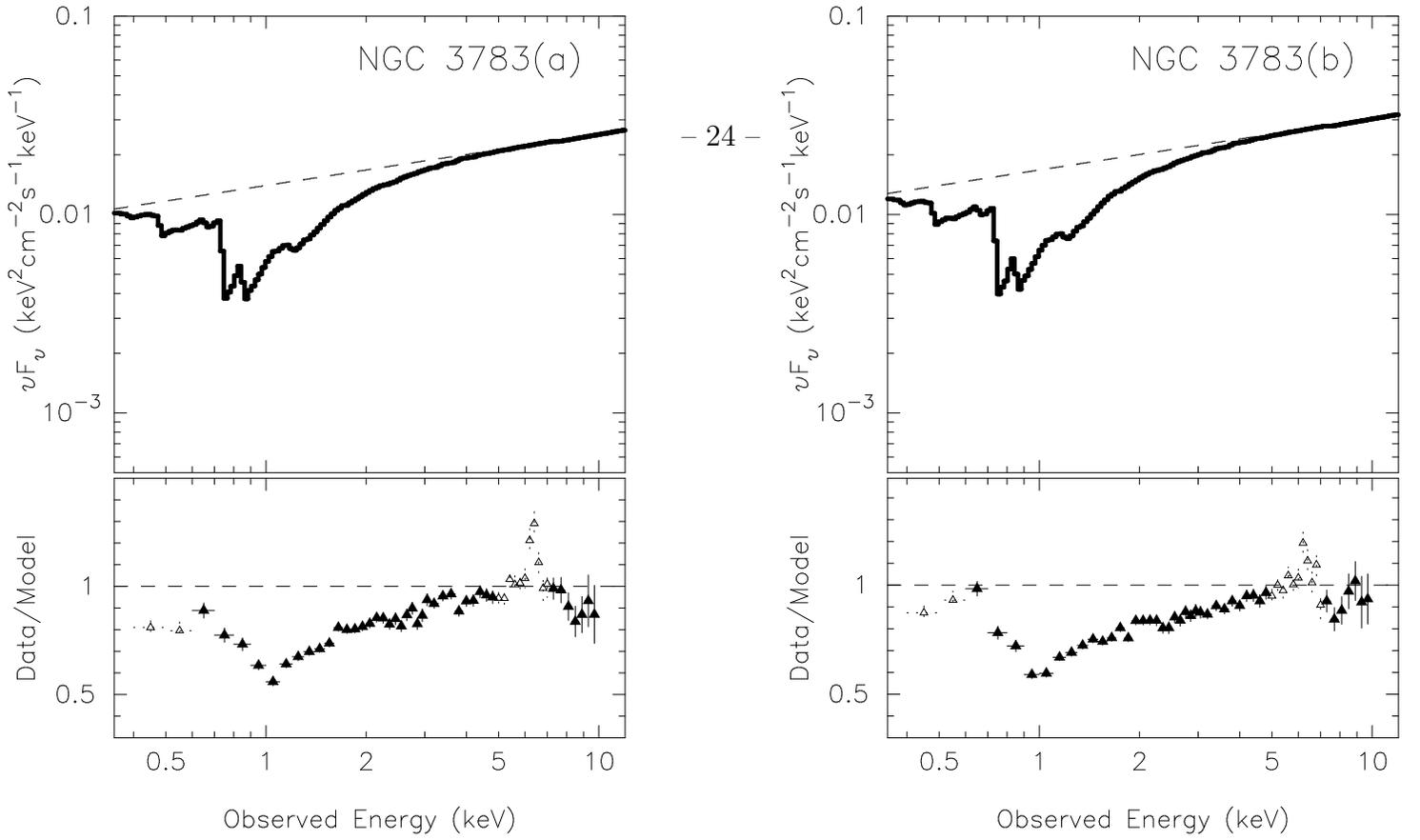}{10cm}{0}{50}{50}{10}{-200}
\caption{The mean data/model ratios residuals for the 1993 observations 
of NGC~3783 
(with symbols as for Fig.~\ref{fig:all96_ion08_ngal_noline_ufmodelratio})
assuming an underlying photon index $< \Gamma >_{96} = 1.73$, 
a screen of ionized material within the column--of--sight of column density
$< N_{H,z} >_{96} = 1.32\times10^{22}\ {\rm cm^{-2}}$
and where the ionization parameter during each observation is given by
(see \S\ref{Sec:abs_only_93})
$U_X = 0.138 (L_{0.5-2}/2.07\times10^{43}\ {\rm erg\ s^{-1}})$.
Excess opacity is clearly evident indicating different gas 
is contributing to the the 1993 observations compared to the 1996 
observations.
\label{fig:ion_ngal_96mean_renorm_reionp}}
\end{figure}

The simplest scenario one can anticipate is that during the 1993 observations
the column--of--sight to NGC~3783 is covered by the same screen of material 
as that present during the 1996 observations (i.e. with $< N_{H,z} >_{96}$), 
but that the level of ionization in this material is lower due to the reduction 
in the ionizing luminosity during the earlier epoch.
However we find this not to be the case.
This is demonstrated in Fig~\ref{fig:ion_ngal_96mean_renorm_reionp}
in which we show the data/model residuals to the 1993 datasets under the 
assumption that the ionized material has $< N_{H,z} >_{96}$, that the 
underlying spectral index is the same at both epochs ($< \Gamma >_{96}$), 
and the ionization parameter during the 1993 observations is lower by a 
factor proportional to that implied by the luminosity of the photoionizing 
continuum.
Since incident photons with energies in the 0.5--2.0~keV dominate the 
ionization structure of highly-ionized species of O, we have rescaled 
$U_X$ assuming $U_X = < U_X >_{96} (L_{0.5-2}/ <L_{0.5-2}>_{96}$)
for Fig~\ref{fig:ion_ngal_96mean_renorm_reionp}.
Large deficits are clearly evident in the residuals, clearly indicating
extra opacity during the 1993 observations. 
Thus we can reject the hypothesis that the same column of gas is in 
equilibrium during the 1993 and 1996 observations with its ionization 
structure simply reflecting the change in ionizing continuum between the 
epochs.

\begin{figure}
\plotfiddle{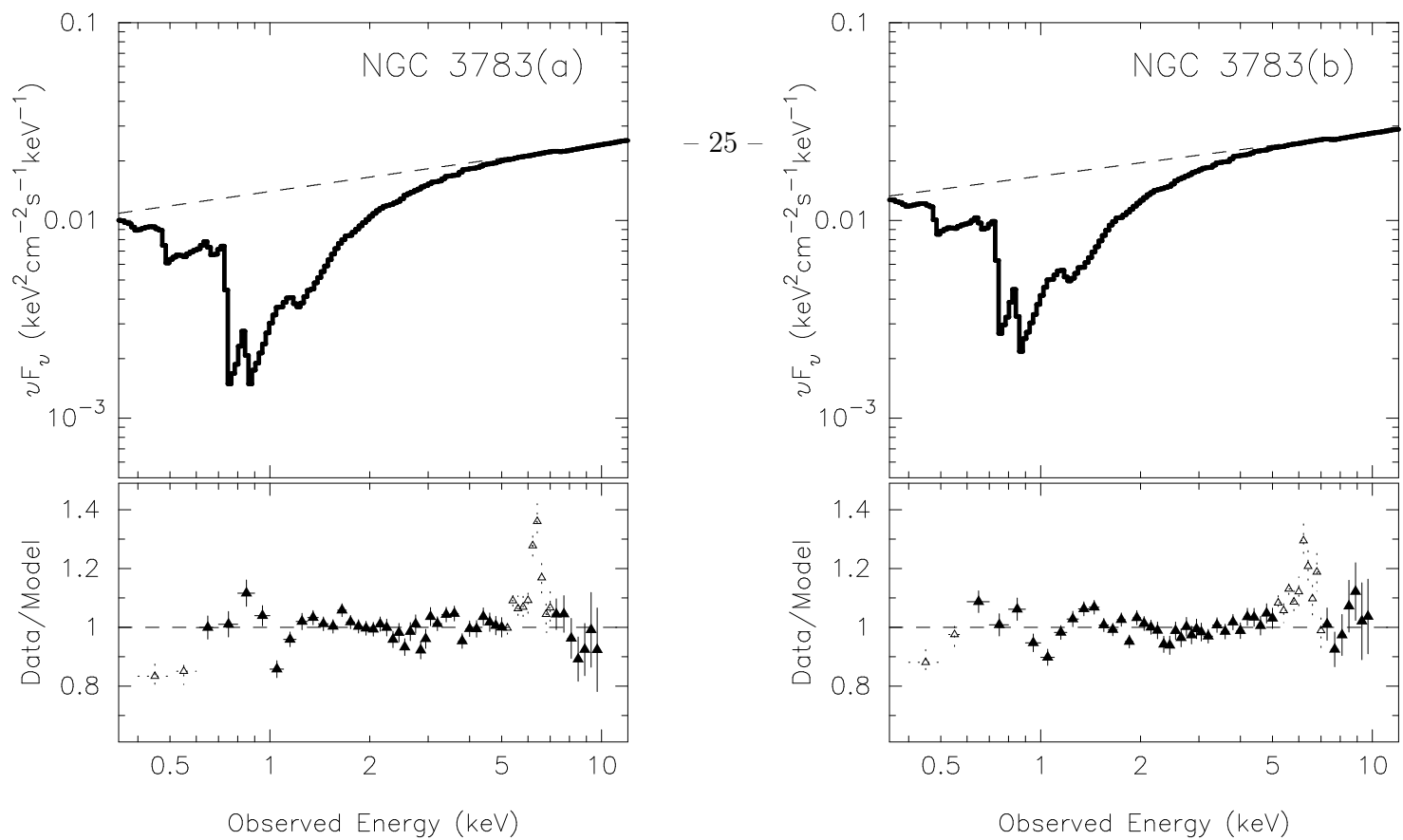}{10cm}{0}{50}{50}{10}{-200}
\caption{As for Fig.~\ref{fig:all96_ion08_ngal_noline_ufmodelratio}, 
showing the best-fitting, single-zone
photoionization model obtained from the spectral
analysis of the individual {\it ASCA} observations obtained during 1993 
(see \S\ref{Sec:abs_only_93}).
It should be noted that despite explaining the bulk of the opacity,
a '1~keV deficit' is clearly present, and is deeper than that 
present in the fits to the 1996 data
(Fig.~\ref{fig:all96_ion08_ngal_noline_ufmodelratio}).
\label{fig:ion_ngal_ufmodelratio_93only}}
\end{figure}

The next simplest scenario is that during both the 1993 and 1996 observations
there is a single zone, in equilibrium, fully covering the cylinder--of--sight,
however the column density, density and/or radius of this material is different
during the two epochs. 
This might be the case if, for example, the circumnuclear material is 
inhomogeneous on large scales (greater than the size of the central source) 
and exhibits transverse motion. 
A different screen of material might then lie within the cylinder--of--sight 
during the two epochs.
However, from Table~3, it can be seen that the single-zone
photoionization model does not provide a statistically acceptable description 
to the 2 datasets obtained during the 1993 epoch.
The spectra implied by these observations are shown in 
Fig.~\ref{fig:ion_ngal_ufmodelratio_93only}, along with the 
mean data/model ratios.
We consider the inadequacy of the single-zone fits an indication that the 
true physical situation is yet more complex.

\subsection{Resolving the discrepancy}
\label{Sec:93vs96}

We observed that in NGC~3783, the opacity of the warm-absorber decreased 
between the 1993 and 1996 epochs, while the source luminosity increased. 
We have also observed a deficit in the data at $\sim$1~keV during both 
epochs compared to the expectations of our ionized-absorber model.
Furthermore, comparison of Figs.~\ref{fig:all96_ion08_ngal_noline_ufmodelratio}
and \ref{fig:ion_ngal_ufmodelratio_93only} reveals that this deficit appears 
more pronounced during the 1993 observations when the implied column density 
of ionized material is larger. 
In this section we discuss several possible explanations for these 
observable properties. 

\subsubsection{Is the 1~keV Deficit an Instrumental Artifact ?}

There are no {\it known} problems with the calibration of the instruments at
$\sim$1~keV during the epochs of these observations (Yaqoob, p.comm). In
addition, as noted in G98, the 1~keV deficit appear to be seen only in the
sources with strong absorption features due to ionized gas (e.g. NGC~3516,
MCG-6-30-15), and yet not seen in sources with no evidence for absorption by
ionized gas (e.g. 3C~120). Thus, whilst a currently-unknown instrumental
artifact cannot be completely excluded as the cause of the feature, such an
explanation requires an unlikely conspiracy of a time-dependent phenomenon.

\subsubsection{Could the Atomic Cross-sections be Inaccurate ?}

The absorption cross-sections used here were based on the analytical fits 
to the theoretical calculations of Verner \& Yakovlev (1995).
These generally agree very well with other calculations and measurements,
giving threshold cross-sections of
$2.4\times10^{-19}\ {\rm cm^2}$ and $9.9\times10^{-20}\ {\rm cm^2}$
for O{\sc vii}(739~eV) and  O{\sc viii}(871~eV), respectively.
However, if these values were in error
(e.g. if the assumed cross-section for O{\sc viii}(871~eV) has 
been significantly underestimated whilst that for O{\sc vii}(739~eV) is 
correct) this might lead to deficit similar to that observed.

In order to investigate this further, we have parameterized the 
1~keV deficit using an additional absorption edge with a threshold fixed 
at a rest-frame energy of 871~eV.
The results are listed in Table~4.
It can be seen that both the 1993 datasets give a best-fitting
column of $N_{H,z} = 1.7 \times 10^{22} {\rm cm}^{-2}$ and 
$\tau \sim 0.5$, while the analysis of the combined 1996 datasets
give a best-fitting
column of $N_{H,z} = 1.1 \times 10^{22} {\rm cm}^{-2}$ and
$\tau \sim 0.15$.
Thus the cross-section of O{\sc viii}(871~eV) would have to 
have been underestimated by a factor of $\sim$2 or 3
(or the cross-section of O{\sc vii}(739~eV) overestimated 
by these factors)
in order to account for the size of the observed deficit.
We consider such an explanation unlikely.

\subsubsection{Is the gas out of equilibrium or turbulent ?}
\label{Sec:nonequilibrium}

Our {\tt ION} models assume that the ionized material is in photoionization 
equibilibrium.
If this is not the case, then we might expect imperfect spectral fits and
deviations of the data compared to the best-fitting model.
The fact that the 1~keV deficit appears more marked in the 1993 observations
would suggest the gas might be further from equilibrium during that epoch.
Furthermore we note that the underlying continuum exhibited marked downward 
trends during both the 1993 observations (Fig.~\ref{fig:lc5760}), which
might be considered circumstantial evidence for such an hypothesis.

In the absence of other sources of heating and cooling, the gas will be in 
photoionization equilibrium if the photoionization and recombination 
timescales of the dominant species are shorter than the variability timescale 
($t_{var}$) of the illuminating continuum.
To first order $t_{ion} \simeq N_{H,z} /U_X n c\ {\rm s}$. 
Thus substituting our best-fitting values to the 1996 observations 
($< N_{H,z} >_{96}$ and $<U_X>_{96}$), $t_{var} > t_{ion}$ if 
$ n \gtrsim 3\times10^{12} t_{var}^{-1}\ {\rm cm^{-3}}$.
The recombination timescale for a given ionic species 
is given by $t_{rec} = 1/\alpha_r(T) n$, 
where $\alpha_r(T)$ is the recombination coefficient for 
gas at an equilibrium temperature $T$.
Assuming $\alpha_r(T)$ from Verner \& Ferland (1996), 
for a given value of $T$, 
$t_{rec}$(O{\sc viii}) is a factor $\sim$2 smaller than
$t_{rec}$(O{\sc vii}), and hence we 
require
$ n \gtrsim 3\times10^{8} T^{0.5} t_{var}^{-1}\ {\rm cm^{-3}}$.
Thus, for $t_{var} = 5\times10^{4}\ {\rm s}$ (the approximate 
timescale in which factor $\sim$2 changes in flux are observed,
Fig.~\ref{fig:lc5760})
and $T \sim 10^5\ {\rm K}$,
we require $ n \gtrsim 10^{8} {\rm cm^{-3}}$ for equilibrium, consistent 
with the density assumed in our {\tt ION} models.
Using eqn.\ref{eqn:U_X}, the derived value of
$< L_{0.1-10} >_{96}$ and $< U_X >_{96}$, such densities require
$r \lesssim 20$ light-days.

Our {\tt ION} models also assume that the ionized material has no
significant internal turbulance in excess of that predicted 
by the local sound speed. If such turbulance is in fact present, then 
features due to resonance-scattering may become visible in the spectrum.
With such a scenario in mind we have also considered 
other parameterizations of the 1~keV deficit in the 1993 datasets.
From Table~4 it can be seen that both 
'notch' and gaussian models have energies in the range 0.9--1.1~keV.
There are a number of resonance transitions in this band, the 
strongest of which are due to Ne{\sc ix}--{\sc x} and Fe{\sc xix}--{\sc xxi}
assuming cosmic abundances.
It is possible that the 1~keV deficit is due to resonance-scattering by one 
or more of these species in a turbulant region (velocities $\gtrsim 0.02c$).
However there are a large number of other resonance transitions throughout 
the 0.6--1.5~keV band. 
For the implied level of ionization, strong features would also be 
expected elsewhere (especially in the 0.6--0.9~keV band due to 
O{\sc vii}--{\sc viii}).
Given the resolution of {\it ASCA}, certain juxtapositions of such features 
can mimic a series of bound-free edges.
Unfortunately it is not possible to determine what fraction of the 
opacity (if any) is due to resonance-scattering using the 
the current data.

Thus non-equilibrium and/or turbulent effects cannot be excluded, but
we suggest they are unlikely to be solely responsible for the 
observed deficit.

\subsubsection{Is the emission from the gas important ?}
\label{Sec:ion_emis}

The ionized material responsible for the absorption features will also give 
rise to emission features due to bound--bound and recombination transitions. 
In addition, electrons in the gas will scatter the continuum. 
Hence if the material subtends a significant solid angle ($\Omega$) as seen 
from the central source, such components may be visible in the observed 
spectrum.
For the values of $N_{H,z}$ and $U_{X}$ implied in the case of NGC~3783, 
the most prominent of these features within the {\it ASCA} bandpass are the 
O{\sc vii}(568--574~eV) and O{\sc viii}(654~eV) emission lines 
(e.g. Netzer 1996). 
We have repeated the spectral analysis assuming an ionized absorber and the 
spectrum expected from the corresponding 'photoionized-emitter'. 
The values of $N_{H,z}$ and $U_{X}$ of the emitting gas were fixed to be the 
same as for the absorbing material (hence giving only one additional 
parameter, $\Omega$, compared to the model used for 
Table~3).

In the case of the 1996 datasets, we find a negligible improvement in the 
quality of the fit compared to the model without the ionized--emitter 
($\Delta \chi^{2} \lesssim 2$ in all cases).
At 90\% confidence (for 1 interesting parameter) we find 
$\Omega/4\pi \leq 0.48$.
Significant improvements in the quality of the fits
are obtained by including the ionized--emitter in the 1993 datasets
(at $>99$\% confidence using the $F$-test in the case of Obs.a).
We find the allowed range in $\Omega/4\pi$ to be 
0.6--1.0 and 0.3--0.8 (at 68\% confidence for 4 interesting parameters)
for Obs.a and b (respectively).
However, in the case of Obs.a, the best-fitting model introduces 
positive residuals in the data/model ratio in the 0.5--1~keV band,
not previously present.
Thus we cannot explain the 1 keV deficit without introducing other 
deviations between model and data. Thus we 
do not consider this a satisfactory explanation of the discrepancies 
between the data and models.

\subsubsection{Is the absorber patchy ?}

If the screen of ionized material is patchy, then the observer may 
have lines-of-sight to the nucleus which undergo different 
degrees of attenuation.
We tested such an hypothesis by first allowing a fraction of 
the nuclear emission to be observed without any attentuation
(besides $N_{H,0}^{gal}$ due to our galaxy).
If more than a few percent of the nuclear emission escapes absorption, then 
this effect might be detectable in the {\it ASCA}\ data. 
However, this extra degree of freedom produced no improvement to the fits. 
We also tested a model in which a fraction of the continuum 
is attenuated by ionized gas with $U_x$ and $N_{H,z}$ whilst the remainder is 
attenuated by material with a different $U_x$ and/or $N_{H,z}$.
However even this more complex model produced no significant improvement to 
the fits. 
Thus we conclude that the 1 keV deficit is not the artifact of a patchy 
absorber.

\subsubsection{Are there two (or more) zones of ionized gas ?}
\label{Sec:2zone}

Another possibility is that there are two (or more) zones of ionized material
surrounding the nucleus, and that the transmission of radiation through 
one zone and then the next, yields a different spectral shape to 
transmission through a single-zone of gas. 
In this case, the simplest picture is that one of the zones is responsible for 
the bulk of the opacity during the 1996 observations and is also present 
during the 1993 observations.
Thus this zone, with a column density $< N_{H,z} >_{96}$, can be considered 
the 'ambient' gas.
During the 1996 observations, an additional zone (cloud) is responsible 
only for the small 1~keV deficit evident in 
Fig.~\ref{fig:all96_ion08_ngal_noline_ufmodelratio}.
However during the 1993 observations, the additional zone contributes 
significant opacity as demonstrated in 
Fig.~\ref{fig:ion_ngal_96mean_renorm_reionp}.
Thus both zones are required for both epochs, but a significant change is only 
required in one of the zones, to explain the observed variations in opacity.

It is difficult to constrain any two-zone, ionized-absorber using these 
{\it ASCA}\ data, and additional complexity arises since one must consider 
the relative covering fractions, cloud sizes, radii and densities of the two 
zones. 
As a full treatment of the two-zone case is beyond the scope of this paper, 
we make a simple paramaterization of the attenuation in excess of the 
'ambient layer', by adding the expected absorption edges to the 
model for the 1993 data.  
We assume the ambient absorber dominates the 1996 absorption, and fix its 
column density at $< N_{H,z} >_{96} = 1.33\times10^{22}\ {\rm cm^{-2}}$. 
We assume this zone of gas responds simply to changes in the source flux, and 
thus we have calculated and fixed the ionization parameter of the ambient 
absorber for each 1993 dataset assuming 
$U_X = < U_X >_{96} \times (L_{0.5-2} / < L_{0.5-2} >_{96})$
where $< U_X >_{96}$ and $< L_{0.5-2} >_{96}$ are from our analysis above.
The threshold energies of the absorption edges were fixed at the appropriate 
values, and those with some evidence for non-negliable values of $\tau$ are 
listed in Table~5. 
It can be seen that as expected O{\sc viii} is the dominant feature in the 
second absorber.
The depth of this extra edge, $\tau$(O{\sc viii})$\simeq 0.6^{+0.2}_{-0.2}$,
implies a column density in O{\sc viii} 
$\simeq 6^{+2}_{-2}\times10^{18}\ {\rm cm^{-2}}$.
For cosmic abundances, and allowing for the unknown fraction of fully-stripped 
O, this imples an effective hydrogen column density in this second zone
$\gtrsim 10^{23} {\rm cm}^{-2}$.
Excess O{\sc viii} can also explain the smaller 1 keV deficit in the 1996 data, 
in this case $\tau$(O{\sc viii})$\simeq 0.15$ 
(Table~4). 
Thus, in the context of the two zone model, the column in the second zone
dropped by a factor of $\sim$ 4 between the 1993 and 1996 epochs.

\section{SPECTRAL ANALYSIS OF THE IRON $K$-SHELL EMISSION}
\label{Sec:nldl}

It is clear from the data/model residuals shown in 
Figs.~\ref{fig:all96_ion08_ngal_noline_ufmodelratio}, 
\ref{fig:ion_ngal_96mean_renorm_reionp}
and \ref{fig:ion_ngal_ufmodelratio_93only}
that the spectrum of NGC~3783 contains a strong emission component in the 
5--7~keV band at both epochs. 
Such a feature is very common in Seyfert galaxies and interpreted in 
terms of a $K\alpha$ fluorescence line from Fe (e.g. Nandra \& Pounds 1994).
Furthermore, it can be seen that there is some indication that the line is 
asymmetric, skewed to lower energies in at least some of the datasets.
Recent {\it ASCA}\ observations have shown such asymmetries are common in 
Seyfert galaxies (e.g. Nandra et al 1997b; Turner et al 1998, and 
references therein), and have been interpreted to arise as the result of the 
extreme Doppler and gravitational effects suffered by line photons emitted 
close to the putative black hole (Mushotzky et al 1995; Tanaka et al 1995).

As noted above, the best-fitting parameters of the emission line are dependent
upon the form of the underlying continuum. Since the best-fitting
continuum is correlated with the absorption present in the soft X-ray band,
we have repeated the spectral analysis, but now also including the
data within the 5--7~keV band. 
Thus we have taken a model consisting of an underlying powerlaw, an 
ionized-absorber, and have added spectral components to model the Fe emission.
For simplicity we have used the single-zone model for the ionized absorber.
This leads to the re-introduction of the 1~keV deficit, but we believe 
such an approach provides a good estimation of the opacity, and hence 
slope of the continuuum in the 5--7~keV band.

In order to keep our analysis as unbiased as possible, we 
chose to parameterize the Fe $K\alpha$
emission as a combination of 
	the emission expected from the innermost regions of 
		the putative accretion flow
and
	a narrow ($<10$~eV) component appropriate for near-neutral Fe.
We also include the Fe $K\beta$ emission assuming an intensity 10\% of that 
of the corresponding $K\alpha$ component.
The Fe emission from the innermost regions of the accretion flow was 
parameterized following Fabian et al. (1989).
Namely we calculate the profile for a line of rest-frame energy $E_z$
assuming a planar geometry where the inclination of our line-of-sight with 
respect to the normal to the plane is given by $i$, and in which the line 
emissivity as a function of radius ($r$) from a Schwarzschild black hole is 
proportional to $r^{-q}$ over the range $R_i < r < R_o$, and zero elsewhere. 
Below we refer to this component as the 'diskline' emission.
The narrow component was included with rest-energies of 6.4~keV and 7.06~keV 
for Fe $K\alpha$ and $K\beta$ (respectively) to allow for the possibility of 
other structures in the circumnuclear environment of the AGN. 
Such structures (for example the outer regions of the accretion flow, the 
putative molecular torus, etc) could be sufficiently distant that relativistic 
effects are unimportant and Fe is in a near-neutral state ($<$ Fe{\sc xiii}), 
yet still subtend a sufficient solid angle at the central source to 
contribute to the Fe emission observed.

As in \S\ref{Sec:abs_only}, we first discuss our results from the 1996 
observations, and then compare these to the 1993 observations.

\subsection{The 1996 observations}
\label{Sec:Fe_96}

\begin{figure}
\plotfiddle{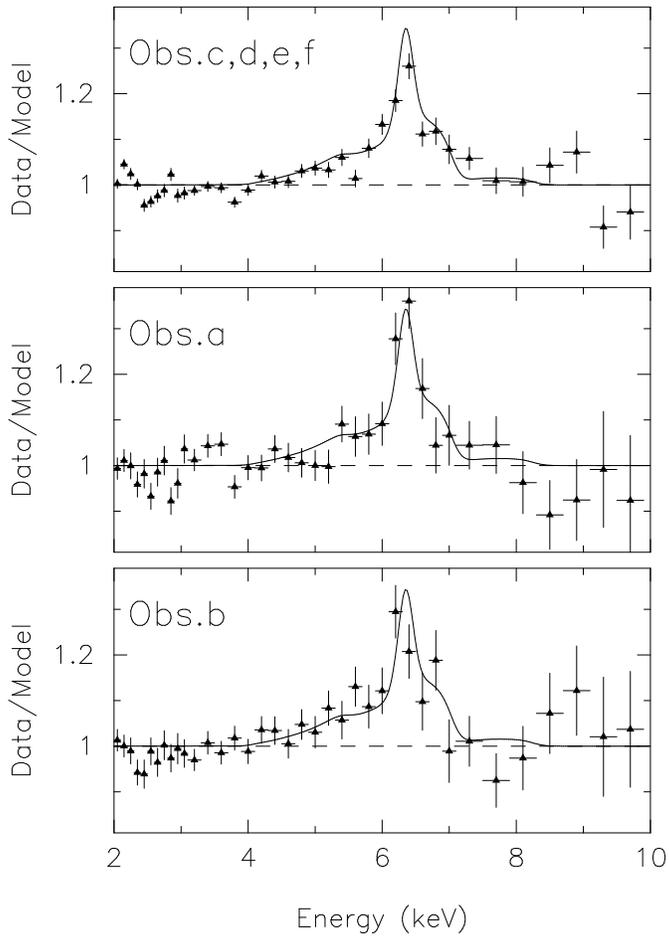}{10cm}{0}{50}{50}{10}{-200}
\caption{The filled points show the mean 
data/model ratios in the Fe $K$-shell regime
assuming the best-fitting models described in 
\S\ref{Sec:abs_only}.
In all three panels
the solid line shows the best-fitting parameterization of the 
Fe emission to the 1996 data
(as discussed in \S\ref{Sec:nldl}) convolved 
with the spectral resonse of the SIS detectors.
It is clear that there is no significant evidence for 
a change in either shape or equivalent width of the Fe emission.
\label{fig:profiles}}
\end{figure}

Unfortunately we are unable to constrain all the parameters associated with 
the diskline component using the individual datasets.
Thus, as above we have chosen to analyse all the datasets obtained during 1996 
simultaneously.
Fixing $R_i$ at $6r_g$ (where $r_g = GM/c^2$ is the gravitational radius 
of the black hole of mass $M$), appropriate for the innermost stable orbit 
within the Schwarzschild metric, we find best-fitting inclination 
$i = 26^{+12}_{-12}$~degrees (at 68\% confidence for 9 interesting parameters). 
The best-fitting value of $q$ pegged at the largest value ($q = 3$), although 
values over the entire range allowed $0 \leq q \leq 3$ are acceptable at 
90\% confidence.
The equivalent widths of the $K\alpha$ lines are 
$EW_{dl} = 230\pm60$~eV and 
$EW_{nl} = 60\pm20$~eV
(at 68\% confidence for 2 interesting parameters).
Fig.\ref{fig:profiles} shows the data/model ratio in the iron K regime, 
illustrating the implied line profile, compared to the model described above. 
Given the steepness of the implied emissivity, $R_o$ could not be constrained 
and we therefore consider it uninteresting.

From Fig.\ref{fig:profiles} it can be seen that there is significant Fe 
$K\alpha$ emission bluewards of $6.4$~keV. 
Indeed the best-fitting value of diskline energy is $E_z = 6.962$~keV for 
the $K\alpha$ component, indicative of Fe{\sc xxvi}.
However the 90\% confidence range for 9 interesting parameters extends down to 
$6.4$~keV. 
In any case, the narrow component of the line could be from highly ionized 
iron, while the diskline is from neutral material. 
As the narrow component is too strong to arise from the warm absorber, the 
former suggestion seems the most compelling.
We find the inclusion of the narrow line component provides a significant 
improvement to the fit ($\Delta \chi^2 = 19$). 
Excluding this component could lead us to the wrong conclusion as to the 
detailed shape of the diskline, thus it is also included below when fitting the 
1993 data. 
Consideration of the Compton-reflection component, with an inclination equal to 
that of the diskline, yields a constraint $\Omega_{R}/2\pi < 1.5$ on the solid 
angle subtended to the continuum source. 

The spectral fits which included this consideration of the Fe $K$-shell 
regime found best-fitting values of $N_{H,z}$, $U_X$ and $\Gamma$
gas as found in \S\ref{Sec:abs_only}.

\subsection{The 1993 observations}
\label{Sec:Fe_93}

In the lower two panels of Fig.\ref{fig:profiles} we compare the 
1993 data with the composite line profile obtained from fitting the 1996 
data. This shows that both the shape and equivalent width of the iron line 
are consistent with the 1996 data. This was confirmed by fits to the 1993 
data.
Again, the parameters associated with the ionized absorber ($N_{H,z}$ and 
$U_X$) were consistent with those found in  \S\ref{Sec:abs_only}.

\section{SUMMARY \& CONCLUSIONS}
\label{Sec:Discuss}

\subsection{The ionized-absorber}
\label{Sec:disc-ionized}

Our observations confirm the presence of substantial opacity due to ionized 
material within the column--of--sight to the X-ray emitting region in NGC~3783. 
Furthermore we have found this opacity in the $\sim 0.7$--1.0~keV band
to decrease by a factor $\sim$2 in the 18 months between the observations 
carried out in 1993 and 1996. 
This large change in opacity is inconsistent with that expected if the 
ionized material is simply responding to changes in the illuminating 
continuum. 
Such behaviour has been seen in at least two other objects 
(MCG-6-30-15, Fabian et al 1994; NGC~3227, Ptak et al 1994, George et al 1998b).

During both epochs we find the bulk of the opacity to be reasonably well 
modelled by a screen of ionized material. 
However, in all cases, we find a deficit is present in the data/model residuals.
In \S\ref{Sec:93vs96} we briefly explored a number of possible 
explanations for such a deficit. 
We suggest that the most attractive explanation is in terms of there being 
two or more zones of photoionized material (\S\ref{Sec:2zone}). 
In this case the column variability can be attributed to a change in column of 
just one of the zones, which might then be inferred to exist as clouds 
moving into and out of the column--of--sight. 
Such an hypothesis has been suggested in other objects based on the 
differential variability in the depth of the O{\sc vii} and O{\sc viii} edges 
(e.g. NGC~3227, Ptak et al 1994; MCG-6-30-15, Otani et al 1996; 
NGC~4051, Guainazzi et al 1996), and/or, as here, spectroscopically based on 
detailed photoionization calculations (e.g. NGC~3516, Kriss et al 1996).
Of course future observations will reveal whether the zone referred to as the 
'ambient gas' in \S\ref{Sec:2zone} truly always lies within the 
column--of--sight. 
A viable alternative is that this zone is in fact simply a slower moving 
(or larger) structure traversing the the column--of--sight.

Absorption features due to C{\sc iv}, N{\sc v} and possibly 
H{\sc i} have been observed in the ultraviolet (UV) spectrum of NGC~3783 
(e.g. Shields \& Hamann 1997, and references therein).
Furthermore there is good evidence that the implied column density 
in these ions is variable (e.g. Maran et al 1996).
Based on the column densities implied by the 1993 {\it ASCA}\ observations of 
the O{\sc vii} and O{\sc viii} edges as given in G95, Shields \& Hamann 
demonstrated that the predicted column densities in the UV absorption lines 
to be in moderately good agreement with the observed range.
Hence it was suggested that there might be a single zone of ionized material 
responsible for the absorption features in both the X-ray and UV 
regimes.
Since the UV absorption due to C{\sc iv} is seen imprinted on the 
corresponding broad emission line, this would clearly place 
the location of the ionized material at a radius larger than the 
region emitting the bulk of the broad C{\sc iv} emission
($\gtrsim $ several light-days).
This is in agreement with our estimate in \S\ref{Sec:nonequilibrium} 
that the material had to be at a radius $\lesssim 20$ light-days in order to 
be in ionization equilibrium.
In Table~6 we list the column densities of the abundant 
Li-like ions predicted from our single-zone model of the ambient gas implied 
by the {\it ASCA}\ observations in 1996.
We find column densities in the strong UV absorption lines within an order of 
magnitude of those observed, confirming the basic conclusion of 
Shields \& Hamann.
However, as acknowledged by Shields \& Hamann and several other workers 
(e.g. Mathur 1997 and references therein), such comparisons are extremely 
sensitive to the form of the photoionizing continuum in the unobserved 
13.6--300~eV band, and the assumed elemental abundances.
Given these facts, and the observed variability, more detailed comparisons 
require simultaneous X-ray and UV observations of a wider species of ions
such as may be possible with the launch of {\it AXAF}, {\it XMM} and {\it FUSE}.

Unfortunately, due to the uncertainities in the calibration of the XRT/SIS 
instrument below 0.6~keV (and hence our exclusion of these data from our 
detailed spectral analysis) our data are unable to provide stringent 
constraints on the intensity of the emission features predicted to arise 
within the ionized material.
However, as noted in \S\ref{Sec:ion_emis}, the limits provided by the current 
data are consistent with the material subtending a solid 
angle $\lesssim2\pi$ at the central source.

\subsection{The Fe emission line}
\label{Sec:disc-line}

Our data confirm the presence of intense Fe $K$-shell emission within NGC~3783, 
with a total eqivalent width ($\sim 300$~eV) and broad, asymetric profile
fairly typical of other Seyfert 1 galaxies (e.g. Nandra et al 
1997b, and references within).
In \S\ref{Sec:nldl} we suggest the line can be deconvolved into two components.
The first is a narrow component with a rest-frame energy of 6.4~keV
(corresponding to an ionization state $\lesssim$Fe{\sc xii}) with an 
equivalent width ($\sim 60$~eV) consistent with production within the distant 
structures within the circumnuclear environment of the central source 
(Ghisellini, Haadt, Matt 1994).
The second component, referred to as the diskline component in 
\S\ref{Sec:nldl}, is that arising close to $6$ gravitational radii of a 
putative supermassive black hole. 
The extreme gravitiational effects suffered by the emission line photons 
at such radii give rise to an asymmetric profile.
We find no compelling evidence that the Fe emission varied in either 
shape or equivalent width between the 1993 and 1996, indicating no
significant changes in the distribution of Fe emissivity.

\acknowledgements

We thank Tahir Yaqoob, Fred Hamann and Joe Shields for useful discussions.
We acknowledge the financial support of the 
Universities Space Research Association (IMG, TJT),
National Research Council (KN), and
the US-Israel Binational Science Foundation and NASA (NH).
This research has made use of 
the Simbad database, operated at CDS, Strasbourg, France; 
and of data obtained through the HEASARC on-line service, provided by
NASA/GSFC.



\clearpage

\clearpage

\clearpage

\clearpage

\clearpage

\clearpage

\clearpage

\clearpage

\end{document}


\ptlandscape
\setlength{\textwidth}{11in}
\setlength{\topmargin}{-0.5in}
\setlength{\oddsidemargin}{-0.7in}
\pagestyle{empty}

\begin{deluxetable}{lrlccccc}

\tablecaption{Observation Log of the {\it ASCA} Observations of 
NGC~3783
\label{tab:obs_log}}

\scriptsize
\tablecolumns{8}
\tablewidth{0pt}
\tablehead{
\colhead{Obs.} &
	\colhead{Seq. No.}	&
	\colhead{Date} &
	\colhead{CCD mode}	&
	\colhead{$t_{exp}$(SIS0)}	&
	\colhead{$Rate$(SIS0)}	&
	\colhead{$N_{pts}$} &
	\colhead{$F_{2-10}$}	
\nl
	&
	&
	&
	&
	\colhead{$(10^{3}\ {\rm s})$}	&
	\colhead{$({\rm count\ s^{-1}})$}	&
	\colhead{(excl 5--7 keV)} & 
	\colhead{$(10^{-12}\ {\rm erg\ cm^{-2}\ s^{-1}})$}	
\nl
	\colhead{(1)}	&
	\colhead{(2)}	&
	\colhead{(3)}	&
	\colhead{(4)}	&
	\colhead{(5)}	&
	\colhead{(6)}	&
	\colhead{(7)}	&
	\colhead{(8)}	
}

\startdata
a	&
	71041000	&
	1993 Dec 19	&
	2	&
	17.3	&
	$0.940\pm0.008$	&
	1038 	&
	\phn$57.7\pm2.9$	
	\nl
b	&
	71041010	&
	1993 Dec 23	&
	2	&
	15.0	&
	$1.117\pm0.009$	&
	1017 	&
	\phn$67.1\pm3.4$	
	\nl
c	&
	74054000	&
	1996 Jul 09 	&
	1	&
	14.9	&
	$1.677\pm0.011$	&
	1091	&
	\phn$88.3\pm4.4$	
	\nl
d	&
	74054010 &
	1996 Jul 11	&
	1	&
	18.7	&
	$1.487\pm0.009$	&
	1098	&
	\phn$80.4\pm4.0$
	\nl
e	&
	74054020	&
	1996 Jul 14 	&
	1	&
	15.5	&
	$1.671\pm0.011$	&
	1082	&
	\phn$86.0\pm4.3$
	\nl
f	&
	74054030	&
	1996 Jul 16	&
	1	&
	16.6	&
	$1.512\pm0.010$	&
	1089	&
	\phn$80.7\pm4.0$
	\nl
\tablecomments{
Col.(1): The observation code used throughout this work.
Col.(2): {\it ASCA} sequence number.
Col.(3): Observation date (start of exposure).
Col.(4): Clocking mode used for the SIS CCDs.
Col.(5): Total exposure time of selected data (for SIS0).
Col.(6): Mean SIS0 count rate in the 0.6--10.0~keV band.
Col.(7): Number of spectral data points (summed over all four instruments)
	in the 0.6--5.0, 7.0--10.0~keV band used in the 
	spectral analysis presented in \S4.
Col.(8): Observed flux in the 2--10~keV band assuming 
	5\% errors.
}
\enddata
\end{deluxetable}

\clearpage

\begin{deluxetable}{lcccc}

\tablecaption{Variability Amplitude ($10^{-2}\sigma^2_{rms}$) for
NGC~3783 
\label{tab:xs_rms_128}}

\scriptsize
\tablecolumns{5}
\tablewidth{0pt}
\tablehead{
\colhead{Obs.} &
	\colhead{SIS Full}	&
	\colhead{SIS Soft}	&
	\colhead{SIS Hard}	&
	\colhead{GIS Hard}	
\nl
	&
	\colhead{(0.5--10~keV)}	&
	\colhead{(0.5--2~keV)}	&
	\colhead{(2--10~keV)}	&
	\colhead{(2--10~keV)}	
\nl
	\colhead{(1)}	&
	\colhead{(2)}	&
	\colhead{(3)}	&
	\colhead{(4)}	&
	\colhead{(5)}	
}

\startdata
a	&
	$0.97\pm0.15$	&
	$1.02\pm0.24$	&
	$0.94\pm0.19$	&
	$0.76\pm0.15$
\nl
b	&
	$0.62\pm0.12$	&
	$0.60\pm0.19$	&
	$0.45\pm0.15$	&
	$0.70\pm0.13$
\nl
c	&
	$0.07\pm0.05$	&
	$0.11\pm0.08$	&
	$0.15\pm0.11$	&
	$0.16\pm0.07$
\nl
d	&
	$0.64\pm0.11$	&
	$1.00\pm0.21$	&
	$0.29\pm0.12$	&
	$0.43\pm0.12$
\nl
e	&
	$0.18\pm0.06$	&
	$0.21\pm0.10$	&
	$0.28\pm0.12$	&
	$0.32\pm0.09$
\nl
f	&
	$0.65\pm0.13$	&
	$0.59\pm0.13$	&
	$0.78\pm0.23$	&
	$0.59\pm0.11$
\nl
\tablecomments{
Col.(1): The observation code.
Cols.(2--5): 
Normalized excess variance, $\sigma^2_{rms}$, and associated error 
calculated using the prescription of Nandra et al (1997a) from light curves 
with a binsize of 128~s.}
\enddata
\end{deluxetable}

\clearpage

\begin{deluxetable}{llll cccclll}

\tablecaption{Results for a single-zone ionized absorber 
(see \S4.2)
\label{tab:ion_ngal_fits}}

\scriptsize
\tablecolumns{11}
\tablewidth{0pt}
\tablehead{
\colhead{Obs} &
	\colhead{$\Gamma$} & 
	\colhead{$N_{H,z}$} &
	\colhead{$U_{X}$} &
	\colhead{$\chi^2$} &
	\colhead{$P$} &
	\colhead{$\frac{\Delta \chi^2_{0.6}}{\Delta N_{0.6}}$} &
	\colhead{$\overline{R_{0.6}}$} &
	\colhead{$L_{0.1-10}$} 	&
	\colhead{$L_{0.5-2}$}  &	
	\colhead{$L_{2-10}$}  
\nl
	&
	&
	\colhead{$(10^{22}\ {\rm cm^{-2}})$}	&
	&
	&
	&
	&
	&
	\colhead{($10^{43}\ {\rm erg\ s^{-1}}$)}	&
	\colhead{($10^{43}\ {\rm erg\ s^{-1}}$)}	&
	\colhead{($10^{43}\ {\rm erg\ s^{-1}}$)}	
\nl
\colhead{(1)}	&
	\colhead{(2)}	&
	\colhead{(3)}	&	
	\colhead{(4)}	&
	\colhead{(5)}	&
	\colhead{(6)}	&
	\colhead{(7)}	&
	\colhead{(8)}	&
	\colhead{(9)}	&
	\colhead{(10)}	&
	\colhead{(11)}	
}

\startdata
a	&
	$1.76^{+0.04}_{-0.04}$	&
	\phn$2.29^{+0.14}_{-0.16}$	&
	\phn$0.139^{+0.011}_{-0.008}$	&
	1134	&
	0.99	&
	2.6	&
	\phn0.84&
	$5.25\pm0.04$	&
	$1.50\pm0.01$	&
	$2.53\pm0.02$
	\nl
b	&
	$1.78^{+0.04}_{-0.04}$	&
	\phn$2.34^{+0.20}_{-0.16}$	&
	\phn$0.167^{+0.006}_{-0.016}$	&
	1113	&
	0.99	&
	1.8		&
	\phn0.93&
	$6.15\pm0.04$	&
	$1.77\pm0.01$	&
	$2.94\pm0.02$
	\nl
c	&
	$1.70^{+0.04}_{-0.03}$	&
	\phn$1.23^{+0.12}_{-0.15}$	&
	\phn$0.140^{+0.012}_{-0.013}$	&
	1111	&
	0.72	&
	2.1	&
	\phn0.99&
	$7.26\pm0.05$	&
	$2.03\pm0.01$	&
	$3.70\pm0.04$
	\nl
d	&
	$1.73^{+0.03}_{-0.04}$	&
	\phn$1.38^{+0.14}_{-0.12}$	&
	\phn$0.139^{+0.011}_{-0.011}$	&
	1113	&
	0.68	&
	5.3	&
	\phn0.82&
	$6.84\pm0.05$	&
	$1.94\pm0.02$	&
	$3.40\pm0.04$
	\nl
e	&
	$1.77^{+0.03}_{-0.04}$	&
	\phn$1.35^{+0.12}_{-0.14}$	&
	\phn$0.138^{+0.012}_{-0.011}$	&
	1153	&
	0.95	&
	2.9		&
	\phn0.92&
	$7.66\pm0.05$	&
	$2.21\pm0.01$	&
	$3.64\pm0.04$
	\nl
f	&
	$1.74^{+0.03}_{-0.03}$	&
	\phn$1.38^{+0.12}_{-0.14}$	&
	\phn$0.137^{+0.012}_{-0.011}$	&
	1127	&
	0.83	&
	2.6		&
	\phn0.93&
	$6.95\pm0.05$	&
	$1.98\pm0.01$	&
	$3.42\pm0.04$
	\nl
\multicolumn{8}{l}{} \nl
\multicolumn{8}{l}{\it weighted-mean values}
\nl
a,b,c,d,e,f	&
	$1.75\pm0.01$		&
	$1.55\pm0.05\ \dag$	&
	$0.151\pm0.004\ \ddag$		&
	\nodata	&
	\nodata	&
	2.9	&
	0.91	&
	$6.49\pm0.01\ \dag$	&
	$1.90\pm0.01\ \dag$	&
	$3.12\pm0.01\ \dag$
\nl
a,b	&
	$1.77\pm0.03$	&
	$2.31\pm0.11$	&
	$0.161\pm0.005\ \ddag$	&
	\nodata	&
	\nodata	&
	2.2	&
	0.89	&
	$5.70\pm0.02\ \dag$ 	&
	$1.64\pm0.01\ \dag$ 	&
	$2.74\pm0.01\ \dag$ 
\nl
c,d,e,f	&
	$1.74\pm0.02$	&
	$1.33\pm0.06$	&
	$0.139\pm0.006$	&
	\nodata	&
	\nodata	&
	3.2	&
	0.92	&
	$7.18\pm0.02\ \dag$ 	&
	$2.07\pm0.01\ \dag$ 	&
	$3.55\pm0.02\ \dag$ 
\nl

\tablecomments{
Results from the fits undertaken in the 0.6--10.0~keV band ({\it excluding}
the 5.0--7.0~keV band) assuming a powerlaw continuum absorbed by 
an ionized absorber at the redshift of the source, plus
a neutral absorber with a column density 
fixed at the Galactic value ($N_{H,0}^{gal} = 8.7\times10^{20}\ {\rm cm^{-2}}$).
The errors on the fit parameters are 68 per cent confidence limits 
for three interesting parameters.
The mean values are calculated weighting the individual values using the 
reciprocal of the corresponding statistial error, with the variance on the 
mean calculated by the reciprocal of the sum of the weights.
$\dag$ best-fitting values inconsistent with a constant at $>$99\% 
confidence.
$\ddag$ best-fitting values inconsistent with a constant at $>$95\% 
confidence.
Col.(1): The observation code(s).
Col.(2): Photon index of the underlying continuum. 
Col.(3--4): Column density ($N_{H,z}$) and ionization parameter ($U_X$) 
of the ionized absorber.
Col.(5): $\chi^2$-statistic for the fit for the $N_{pts}$ data points
	given in Table~\ref{tab:obs_log}.
Col.(6): Probability that the $\chi^2$-statistic should be less than
	the observed value for the number of degrees of freedom, $dof$, 
	given by $N_{pts} - 7$.
Col.(7): Ratio of the increase in the $\chi^2$-statistic to the number of 
	addition data points when the best-fitting model is extrapolated 
	below 0.6~keV.
Col.(8): Mean data/model ratio for data points below 0.6~keV.
Cols.(9--11): Implied luminosity of the underlying continuum 
	(excluding the Fe $K$-shell emission, and corrected 
	for {\it all} absorption along the line-of-sight)
	over the 0.1--10, 0.5--2.0 and 2.0-10~keV energy ranges 
	(respectively)
	assuming 
	$H_0 = 50\ {\rm km\ s^{-1}\ Mpc^{-1}}$
	and $q_0 = 0.5$.
}

\enddata
\end{deluxetable}

\clearpage

\begin{deluxetable}{lllllll cccc}

\tablecaption{Results from parameterizations of the '1~keV deficit'
(see \S4.3)
\label{tab:ion_ngal_o8_fits}}

\scriptsize
\tablecolumns{11}
\tablewidth{0pt}
\tablehead{
\colhead{Obs} &
	\colhead{$\Gamma$} & 
	\colhead{$N_{H,z}$} &
	\colhead{$U_{X}$} &
	\colhead{$E_{z}$} &
	\colhead{p1} &
	\colhead{p2} &
	\colhead{$\chi^2$} &
	\colhead{$P$} &
	\colhead{$\frac{\Delta \chi^2_{0.6}}{\Delta N_{0.6}}$} &
	\colhead{$\overline{R_{0.6}}$} 
\nl
	&
	&
	\colhead{$(10^{22}\ {\rm cm^{-2}})$}	&
	&
	&
	&
	&
\nl
\colhead{(1)}	&
	\colhead{(2)}	&
	\colhead{(3)}	&	
	\colhead{(4)}	&
	\colhead{(5)}	&
	\colhead{(6)}	&
	\colhead{(7)}	&
	\colhead{(8)}	&
	\colhead{(9)}	&
	\colhead{(10)}	&
	\colhead{(11)}
}

\startdata
\multicolumn{11}{l}{} \nl
\multicolumn{11}{l}{{\it edge model} (p1=$\tau$)}
\nl
a	&
	$1.70^{+0.05}_{-0.05}$	&
	\phn$1.66^{+0.30}_{-0.33}$	&
	\phn$0.119^{+0.016}_{-0.018}$	&
	$0.871\ (f)$	&
	$0.51^{+0.23}_{-0.21}$	&
	\nodata	&
	1110	&
	0.96	&
	1.2	&
	\phn0.91	
	\nl
b	&
	$1.72^{+0.05}_{-0.06}$	&
	\phn$1.66^{+0.37}_{-0.34}$	&
	\phn$0.139^{+0.020}_{-0.018}$	&
	$0.871\ (f)$	&
	$0.47^{+0.21}_{-0.25}$	&
	\nodata	&
	1086	&
	0.96	&
	1.6		&
	\phn1.03	
\nl
c,d,e,f	&
	$1.71^{+0.02}_{-0.02}$	&
	\phn$1.10^{+0.15}_{-0.12}$	&
	\phn$0.128^{+0.009}_{-0.009}$	&
	$0.871\ (f)$	&
	$0.15^{+0.07}_{-0.08}$	&
	\nodata	&
	4542	&
	0.98	&
	2.8		&
	\phn0.94	
\nl
\multicolumn{11}{l}{} \nl
\multicolumn{11}{l}{{\it notch model} (p1=$\sigma_n$)}
\nl
a	&
	$1.73^{+0.05}_{-0.05}$	&
	\phn$2.19^{+0.21}_{-0.19}$	&
	\phn$0.144^{+0.015}_{-0.012}$	&
	$1.06^{+0.02}_{-0.02}$	&
	$0.03^{+0.01}_{-0.01}$	&
	\nodata	&
	1091	&
	0.92	&
	2.2	&
	\phn0.86	
\nl
b	&
	$1.76^{+0.05}_{-0.05}$	&
	\phn$2.29^{+0.23}_{-0.25}$	&
	\phn$0.164^{+0.014}_{-0.014}$	&
	$1.07^{+0.04}_{-0.05}$	&
	$0.02^{+0.01}_{-0.01}$	&
	\nodata	&
	1088	&
	0.97	&
	1.6	&
	\phn0.95	
\nl
\multicolumn{11}{l}{} \nl
\multicolumn{11}{l}{{\it gaussian model} (p1=$\sigma_g$, p2=$EW_{g}$)}
\nl
a	&
	$1.72^{+0.06}_{-0.06}$	&
	\phn$2.19^{+0.21}_{-0.25}$	&
	\phn$0.145^{+0.016}_{-0.013}$	&
	$1.07^{+0.03}_{-0.04}$	&
	$0.05^{+0.08}_{-0.05\ (p)}$	&
	$40^{+2}_{-2}$	&
	1090	&
	0.91	&
	2.0	&
	\phn0.86	
\nl
b	&
	$1.73^{+0.06}_{-0.06}$	&
	\phn$2.24^{+0.58}_{-0.26}$	&
	\phn$0.174^{+0.050}_{-0.020}$	&
	$0.99^{+0.09}_{-0.14}$	&
	$0.12^{+0.10}_{-0.08}$	&
	$66^{+4}_{-9}$	&
	1078	&
	0.94	&
	1.4	&
	\phn0.97	
\nl

\tablecomments{
As for Table~\ref{tab:ion_ngal_fits}, but with the inclusion of an 
additional functional form to model the '1~keV deficit'.
The first is an absorption edge, assuming
a transmission of the form
$\exp (-\tau (E/E_{z})^{-2.67})$ for rest-frame energies $E \geq E_{z}$,
and unity elsewhere (i.e. containing 2 additional parameters
$E_{z}$ and $\tau$).
The second is a 'notch', assuming
zero transmission over an energy range
$E_{z} - \sigma_n/2 \leq E \leq E_{z} + \sigma_n/2$
and unity elsewhere (i.e. 2 additional parameters).
The third is a gaussian absorption feature, centered at an energy
$E_{z}$, with a width $\sigma_g$, and equivalent width $EW_{g}$
(i.e. 3 additional parameters).
In all cases, $E_{z}$ was constrained to lie within the
range $0.6 \geq  E_{z} \geq 1.362$~keV, corresponding to the
lower cutoff energy in the datasets and the energy of the $K$-shell
edge of Ne{\sc x} (respectively).
The allowed energies therefore cover the energies of bound--free
$K$-shell transitions of
N{\sc vii}, O{\sc v}--{\sc viii}, and all species of Ne,
along with a large number of bound--bound transitions
in these species.
Column headings as for 
Table~\ref{tab:ion_ngal_fits}, except
cols.(5--7) which give
the rest-frame energy ($E_z$) and additional parameters of the function
	modeling the 1~keV deficit.
}

\enddata
\end{deluxetable}

\clearpage

\begin{deluxetable}{lllll lllll lll l}

\tablecaption{Parameterization of the excess opacity seen during 1993
(see \S4.3.5)
\label{tab:ion96_7smedge00_fits}}

\scriptsize
\tablecolumns{14}
\tablewidth{0pt}
\tablehead{
\colhead{Obs} &
	\colhead{$\Gamma$} & 
	\colhead{$\tau$(O{\sc viii})}	&
	\colhead{$\tau$(Ne{\sc viii})}	&
	\colhead{$\tau$(Ne{\sc ix})}	&
	\colhead{$\tau$(Ne{\sc x})}	&
	\colhead{$\tau$(Fe)}   &
	\colhead{$\chi^2$} &
	\colhead{$P$} &
	\colhead{$\frac{\Delta \chi^2_{0.6}}{\Delta N_{0.6}}$} &
	\colhead{$\overline{R_{0.6}}$} &
	\colhead{$L_{0.1-10}$} 	&
	\colhead{$L_{0.5-2}$}  &	
	\colhead{$L_{2-10}$}  
\nl
\colhead{(1)}	&
	\colhead{(2)}	&
	\colhead{(3)}	&	
	\colhead{(4)}	&
	\colhead{(5)}	&
	\colhead{(6)}	&
	\colhead{(7)}	&
	\colhead{(8)}	&
	\colhead{(9)}	&
	\colhead{(10)}	&
	\colhead{(11)}	&
	\colhead{(12)}	&
	\colhead{(13)}  &
	\colhead{(14)}  
}

\startdata
a	&
	$1.69^{+0.06}_{-0.05}$	&
	$0.61^{+0.17}_{-0.19}$	&
	$0.05^{+0.13}_{-0.05\ (p)}$	&
	$0.02^{+0.14}_{-0.02\ (p)}$	&
	$0.04^{+0.08}_{-0.04\ (p)}$	&
	$0.17^{+0.22}_{-0.17\ (p)}$ &
	1103	&
	0.95	&
	0.7	&
	0.98	&
	$4.84\pm0.03$	&
	$1.36\pm0.01$	&
	$2.51\pm0.02$
\nl
b	&
	$1.73^{+0.05}_{-0.04}$	&
	$0.62^{+0.14}_{-0.14}$	&
	$0.00^{+0.10}_{-0.00\ (p)}$	&
	$0.00^{+0.11}_{-0.00\ (p)}$	&
	$0.09^{+0.06}_{-0.07}$	&
	$0.03^{+0.15}_{-0.03\ (p)}$  &
	1091	&
	0.97	&
	1.7	&
	1.04	&
	$5.82\pm0.03$	&
	$1.65\pm0.01$	&
	$2.87\pm0.02$
\nl

\tablecomments{
Results from the model discussed in \S4.3.5 in which a series of 
absorption edges are added to the opacity expected from the 1996 results.
The ionization parameter was fixed at $U_X = 0.093$ \& $0.112$ for 
Obs.a \& b respectively. 
In both cases $N_{H,z}$ is fixed at $1.32\times10^{22}\ {\rm cm^{-2}}$.
Threshold energies of 	0.871, 1.125, 1.125 \& 1.362~keV were fixed 
for O{\sc viii}, Ne{\sc viii}, Ne{\sc ix} \& Ne{\sc x} respectively.
The threshold energy for Fe were left free over the range 
7.1--9.3~keV, and could not be constrained.
The errors on the fit parameters are 68 per cent confidence limits 
for 7 interesting parameters.
$p$ indicates the parameter 'pegged' at the specified limit.
Column headings as for 
Table~\ref{tab:ion_ngal_fits}, except
cols.(3--7) which give
the optical depth at the threshold energy of each edge.
}

\enddata
\end{deluxetable}

\clearpage

\begin{deluxetable}{lllll}

\tablecaption{Column Densities of Individual Ions 
from the 1996 Observations
\label{tab:coldens}}

\scriptsize
\tablecolumns{5}
\tablewidth{0pt}
\tablehead{
	\colhead{element} &
	\colhead{$A$/H} &
	\multicolumn{3}{c}{$N_{ion}({\rm cm^{-2}})$}	
\nl
	&
	\colhead{$(10^{-4})$}	&
	\colhead{Li-like}	&
	\colhead{He-like}	&
	\colhead{H-like}	
\nl
\colhead{(1)}	&
	\colhead{(2)}	&
	\colhead{(3)}	&	
	\colhead{(4)}	&
	\colhead{(5)}	
}

\startdata
C 	&	$3.63$
	&	$2.0\times10^{13}$	
	&	$8.1\times10^{16}$
	&	$1.3\times10^{18}$
\nl
N	&	$1.12$
	&	$4.0\times10^{14}$      
	&	$1.3\times10^{17}$
	&	$6.8\times10^{17}$
\nl
O	&	$8.51$
	&	$5.9\times10^{16}$      
	&	$3.0\times10^{18}$
	&	$5.3\times10^{18}$
\nl
Ne	&	$1.23$
	&	$1.4\times10^{17}$      
	&	$8.6\times10^{17}$
	&	$3.8\times10^{17}$
\nl
Mg	&	$0.38$
	&	$1.5\times10^{17}$      
	&	$2.5\times10^{17}$
	&	$3.9\times10^{16}$
\nl
Si	&	$0.36$
	&	$1.5\times10^{17}$      
	&	$9.3\times10^{16}$
	&	$5.6\times10^{15}$
\nl
S	&	$0.16$
	&	$3.1\times10^{16}$      
	&	$9.1\times10^{15}$
	&	$2.4\times10^{14}$
\nl
Ar	&	$0.04$
	&	$1.5\times10^{15}$	
	&	$2.7\times10^{14}$
	&	$3.5\times10^{12}$

\tablecomments{
These predictions are for $U_{X}=0.139$,
$N_{H,z}= 1.32\times10^{22}\ {\rm cm^{-2}}$ and hence only for the 
'ambient' material discussed in \S4.3.5.
It should be stressed that such values are very sensitive to the 
form assumed 
for the ionizing continuum in the XUV band (see text),
and of course the elemental abundances.
Col.(1): Atomic element.
Col.(2): Assumed abundance relative to hydrogen.
Cols.(3--5): Derived column density in the specified ion.
}
\enddata
\end{deluxetable}
\clearpage